***Original Research Article***

# Neuroqueering Physics Literacy

**Liam G. McDermott**[1]
*University of Connecticut*

*ABSTRACT:* Neurodivergent people experience the world, conceptualize scientific concepts, and engage with meaning-making differently than their neurotypical peers. When these differences are actualized in an institutional setting, such as higher education, conflict can arise which has a disabling effect on them. One such way this conflict can arise is through defining and subsequent assessment of scientific literacy, as we define scientific literacy through a neurotypical lens. There is much literature on how to better teach neurodivergent students to succeed in a neurotypical-normative scientific environment, however very little research exists which examines what neurodivergent differences in learning look like from these students' perspectives. I here present a case study of five neurodivergent physicists as a part of a broader study on neurodiversity in physics, explaining why it is imperative to redefine scientific literacy and reorient it to a neuroqueer standpoint.

Recent studies have revealed disparities in physics literacy among marginalized groups, highlighting the subjectivity of physics performed by humans (Cochran et al., 2021). Many neurodivergent (e.g. autistic, ADHD, dyslexic) physics students ground their identity in learning and performing physics non-normatively (McDermott, 2022). Research has focused on methods to increase physics literacy, including computational literacy, teaching strategies, and supplementary instruction (Odden et al., 2019; Yuliati et al., 2018; Holmes et al., 2017). However, the way we define and assess "physics literacy" may disenfranchise learners who deviate from neurotypical-normative performance of physics tasks. Discussions on reading-specific literacy have evolved to consider various social factors, such as neuroqueer literacies, which highlight how society's definition of literacy can negatively impact neurodivergent individuals (Kleekamp, 2020; Smilges, 2021). I posit that the discipline-based education research (DBER) community would benefit from applying insights from Kleekamp's (2020) and Smilges's (2021) neuroqueer literacies towards scientific learning.

This paper is an examination of what this neurodivergent physics literacy may look like. This paper aims, in part, to describe the experiences of neurodivergent physicists and advocate for redefining scientific knowledge construction towards a neuroqueer literacy. In the following pages, I make the case for questioning, subverting, and redefining how we understand scientific knowledge construction towards a neuroqueer scientific literacy. I do this by employing a queered method of analysis, deviating from traditional case study method (Bhattakarya, 2017), uncovering themes of neuroqueered physics performance through narrative storytelling.

[1] Liam G. McDermott (he/they) is a postdoctoral scholar at University of Connecticut working with Dr. Erin Scanlon. His research centers physics curriculum development and neurodivergent learning and performance in physics. vtb24001@uconn.edu https://orcid.org/0000-0002-0594-0486

## Key Definitions

***Neurodiversity***–(noun) "The diversity of human minds, the infinite variation in neurocognitive functioning [(ie. thinking, sensing, and behaving)] within our species." (Walker, 2021)

***Neurodivergent Identity***–(noun) An identity characterized by differences in thinking, sensing, and behaving. To be neurodivergent (adjective), then is to identify with these embodied differences in thought, sense, and behavior contra the "norm."

***Neurotypical***–(adjective) Something done in line with "conventional" ways of thinking, sensing, and behaving (Walker, 2014; McDermott, 2022). In that someone identifies as neurotypical, I mean that they identify with conventional ways of thinking, sensing, and behaving.

***Neuroqueer***–(verb) Radical disruption of neurotypical-normative structures, thought, sensing, behavior, and knowledge construction.

## Positionality

I, the author of this paper and person who conducted and analyzed the interviews, am neurodivergent. I claim a disabled identity in multiple ways. I am hard-of-hearing (not a neurodivergent identity), autistic, and have ADHD and OCD. I am also a physics education researcher. I have personally experienced many of the things that participants described in their interviews. My personal experience and identity has aided me in analyzing these interviews and has helped inform my analysis of their statements (e.g.. through mitigating the generalized double empathy problem; Milton, 2012).

As a member of the neurodivergent community, I have a particular understanding of the language complexities of my community (Rodriguez Espinosa et al., 2022). For example, metaphor and sarcasm are communicated differently in neurodivergent vs. neurotypical speech (Hadden, 2023). Study participants used more context-based, rather than tone-based, sarcasm. This is something that I, as a neurodivergent person, am cued in on.

It is also important to note that I am white, and so too are the five participants in this paper. People of Color have historically been excluded from the neurodivergent community due, in part, to medical racism (Ballentine, 2019). Additionally, when conducting interviews, I occupied a very privileged standpoint in which the socially constructed invisibility (Rollock, 2012) of race shaped my perception towards not noticing color, instead seeing a group of people whose experiences as white neurodivergent people reflected my own. I did not notice until after analysis that there was anything weird or wrong about the lack of People of Color in the five narratives. It is imperative when reading this paper, that readers understand that the narratives come from white and privileged perspectives and were analyzed with a white and privileged perspective.

It is also important to note that neurodiversity is itself diverse. The neurodivergent community is a heterogenous one. I conduct analysis from the viewpoint that there is no one way to be neurodivergent, and that setting strict boundaries on what is and is not neuroqueer is not reflective of reality and is counterproductive. I do not know everything about being neurodivergent, but I am open to learning from the experiences of others.

## Neuroqueering Literacies

Developed by Kleekamp (2020) in her work "'No! Turn the Pages!' Repositioning neuroqueer Literacies," and built-on further by Smilges (2021), neuroqueer literacies provides a framework for understanding how neurodivergent people radically disrupt neurotypical-normative reading. Both Kleekamp and Smilges, argue that focusing on a performative model of literacy is ableist. Instead, they argue, we should focus on knowledge construction when we discuss literacy. As an example, Kleekamp (2020) describes a student who uses gestures and an assistive device, rather than pen and paper or speech, to communicate thoughts about the world. This student shows rhetorical competence in making connections between questions asked of him in class by using non-normative means.

Other researchers have drawn attention to neurodivergent differences in knowledge construction such as (1) "rational inattention" (Jones et al., 2023) which posits that some neurodivergent people (specifically those with ADHD) optimize their use of sources of information which quickly resolve misunderstanding, and (2) "neurodivergent intersubjectivity" which posits that some neurodivergent people (specifically autistics) create shared understanding based on assumed common references and a lack of expectation that topics of conversation need to connect (Heasman & Gillespie, 2019). However, none to date have provided a framework for examining scientific knowledge construction. Moreover, these discussions on neurodivergent knowledge construction each focus on one specific neurodivergent identity, and do not necessarily speak to the diversity of the neurodivergent/neuroqueer community. Neuroqueer literacies applied to physics learning can apply to scientific knowledge construction and is broad enough to encompass the diversity present in the neurodivergent community.

Kleekamp (2020) gives three themes when it comes to doing neuroqueer literacies in the classroom. They are:

1. Presuming competence in neuroqueer literacy practices.
2. Asociality as a mode to produce countersocialities in neuroqueer literacy practices.
3. Neuroqueer embodied invention in literacy practices.

Educators should first and foremost assume that students are competent at their assigned tasks, especially if they perform the task contrary to how it is "supposed" to be completed. Moreover, educators should hold space for this counter-performance, in which neurodivergent people invent new ways of knowledge construction.

In fact, when we expand our definition of literacy past Smilges's (2021) "abled-reading," we open ourselves up to an infinite set of possibilities for constructing knowledge. Smilges draws from Johnson and McRuer's (2014) cripistemologies the idea that disabled people create unique knowledge from their embodied identities and perspectives. Thus, to neuroqueer literacy is to ask how the lived experiences and embodied identity of neurodivergent people allows them to construct counter-normative knowledge construction about their environment. Smilges argues that reading-centric literacy is itself disabling for some neurodivergent people. They state that, much like McRuer's (2004) ideas on writing, the rigidity of neurotypical-normative reading and writing serves to disable neurodivergent students, and acts counter to the fluidity and diversity of the human mind. Here, I argue similarly. The rigidity of neurotypical-normative physics learning and performance disables neurodivergent students who create knowledge non-normatively.

Neurotypicality is hegemonic (Radulski, 2022); it is culturally dominant, and power and privilege are given to those who act and think neurotypically. How society operates day to day is thus purposefully constructed from this hegemonic viewpoint. Ableism is normalized in society, intrinsic to everyday life. It is baked-into communication, physical structures, habits, expectations, and etiquette. Ableism is intrinsic to education and the institutions we have built for its purpose.

The skills neurodivergent students show are routinely undervalued or ignored, deemed as inefficient ways to solve problems or simply not useful problem-solving skills (Smilges, 2021).

We see this neurotypical-normative ableism rear its head in how we define success in learning. Students are often assumed to be neurotypical by educators and are taught in a neurotypical way (von Below et al., 2021). Assessment strategies which accompany this style of teaching often are structured to be completed through neurotypical means. Education as an institution is constructed using frameworks which assume neurotypicality is normal and correct (Jurgens, 2020; Wilson, 2017; Jack, 2022; Lang, 2019), so much so that educators can often be empowered to classify students as "metacognitively lacking," proceeding as if their job is to fix these neurodivergent students so that they may act and learn correctly (Jack, 2022, p. 350). Much like the rigidity of neurotypical-normative reading and writing, the rigidity of science classes due to this neurotypical hegemony, and the similar vein of assessment strategies which follow in class, can lead to students being assigned the label of scientifically or field-specifically illiterate. Whether this label comes from content experts, or the students themselves, it is incredibly harmful to identity development as a scientist for students not to be recognized as competent (Hazari et al., 2018; Hyater-Adams et al., 2018).

Neurodivergent scientists neuroqueer scientific knowledge construction due to differences in neurocognitive ability, and differences in sense-making and concept-mapping. However, how we assess these students' scientific ability does not reflect any acknowledgement or appreciation for these differences. I provide the five narratives below to describe the experiences and perspectives of neurodivergent scientists and offer insight into how they neuroqueer their learning.

## Methods

As a part of a broader study examining what it means to be neurodivergent *and* a physicist, I conducted five interviews with neurodivergent physicists. Participants were identified using a convenience sampling method. None of the participants in this paper are past or present students of mine, and participants span multiple institutions and states. The demographic make-up of these participants consists of three graduate students (Marsden, Louis, and Albert) and two undergraduate students (Tom and Elizabeth). Table I breaks down the demographic information of each participant.

| **Participant/Interviewee Pseudonym** | **Identities/Demographics (Gender, Sexuality, Race)** | **Academic Stage** | **Neurodivergent Identity(ies)** |
|---|---|---|---|
| Marsden (she/they) | Nonbinary, asexual, white | Graduate student | OCD, ADHD |
| Louis (he/him) | Cisgender, demisexual, white | Graduate student | BPD |
| Albert (he/him) | Cisgender, bisexual, white | Graduate student | OCD, ADHD |
| Elizabeth (she/her) | Cisgender, bisexual, white | Undergraduate | Autistic, OCD |
| Tom (he/they) | Nonbinary, bisexual, white | Undergraduate | Autistic |

***Table 1.*** *Demographic summary of participants.*

I conducted 1–2-hour semi-structured interviews over Zoom, in which interviewees were asked questions regarding their experience regarding their identity as a neurodivergent person and their identity as a physicist. These interviews consisted of 28 questions which prompted participants to expound on experiences navigating academia with respect to their identity.

During transcription of these interviews, themes emerged in terms of what the interviewees reported in their experience of coming to understand physics. While I analyzed and recounted the interviews, common experiences regarding sense-making, assessment, and science conceptualization kept being repeated. I employed an inductive coding method in my analysis. For the purposes of this paper, I collected participants' quotes which were illustrative of these perceived phenomena of neurodivergent differences of physics learning. These quotes were used to aid in the narrativization of these perspectives, with that narrativization intended to help readers make sense of the data and connect (across difference) with the humanity of these individuals.

Member checking was employed to support this analysis, with my findings and narratives being checked and corroborated by each interviewee to ensure the accuracy of narratives, claims, and my understanding of their experiences. Participants were sent an email by the author containing relevant draft work and prompted to address any concern they had with how they were represented. Participants were given approximately a month to read and review this work, and they provided helpful feedback on their perspectives, thus aiding my analysis. I have no reason to believe that participants felt any pressure that they were required to agree with the findings.

## Five Cases of Neuroqueer Physics Literacy

**Marsden (Graduate student, identifies with OCD, ADHD)**

"You know those fireworks that, like, explode? And then all the little pieces, those explode too?" Marsden asked me using metaphor, as most interviewees did, while discussing how they are different from their peers. They tell me, "that's kind of how I feel like my thoughts work, and I gotta go in a lot of circles to get to the same spot." They continued, clarifying this statement, that they are remarkably pattern-oriented. They shared that oftentimes when they see patterns, as comes up regularly in scientific pursuits, they try to interrogate as many possible patterns as they can simultaneously. Often, this can cause them to "get away from [them]self," losing their train of thought. This happens even when there are not physically relevant patterns in a system, problem, or data, which can result in Marsden taking a longer time to complete tasks for which they do not understand the material presented to an "adequate" level. However, they report that this simultaneous pattern interrogation and "firework thought" process can lead to them developing their own "convoluted" (neutral, from Marsden's own words, comparable to how others might use the word "interconnected") way of understanding science and mathematics. Fig. 1 provides a visualization of this "firework thought" experienced by Marsden and other neurodivergent people.

"[Developing a more convoluted way of understanding] becomes more difficult to do… the more complex the subject is, like the more background you need to understand a subject," Marsden explains, comparing elementary mathematics and their understanding of 9's times tables to doing graduate classical mechanics. They go into more detail, describing their way of understanding a physics problem as "tangles" which they must first untangle to arrive at a solution. They state:

> I have to disentangle situations a lot that other people… the information goes into their brains and it just kind of goes to all the right spots. And then it goes into my brain, and it gets all messed up and twisted around, and I gotta figure out how to untangle it and … if I then, you know, ask… Sometimes when I ask people for help with that, it's like they don't obviously understand the tangles that's going on with the information. And so they like… It can be difficult to figure out what it is that I need to make that information make sense to me.

Contrasting to how neurotypical physicists experience the world, they state that, for their peers, what they struggle with in understanding is "more straightforward."

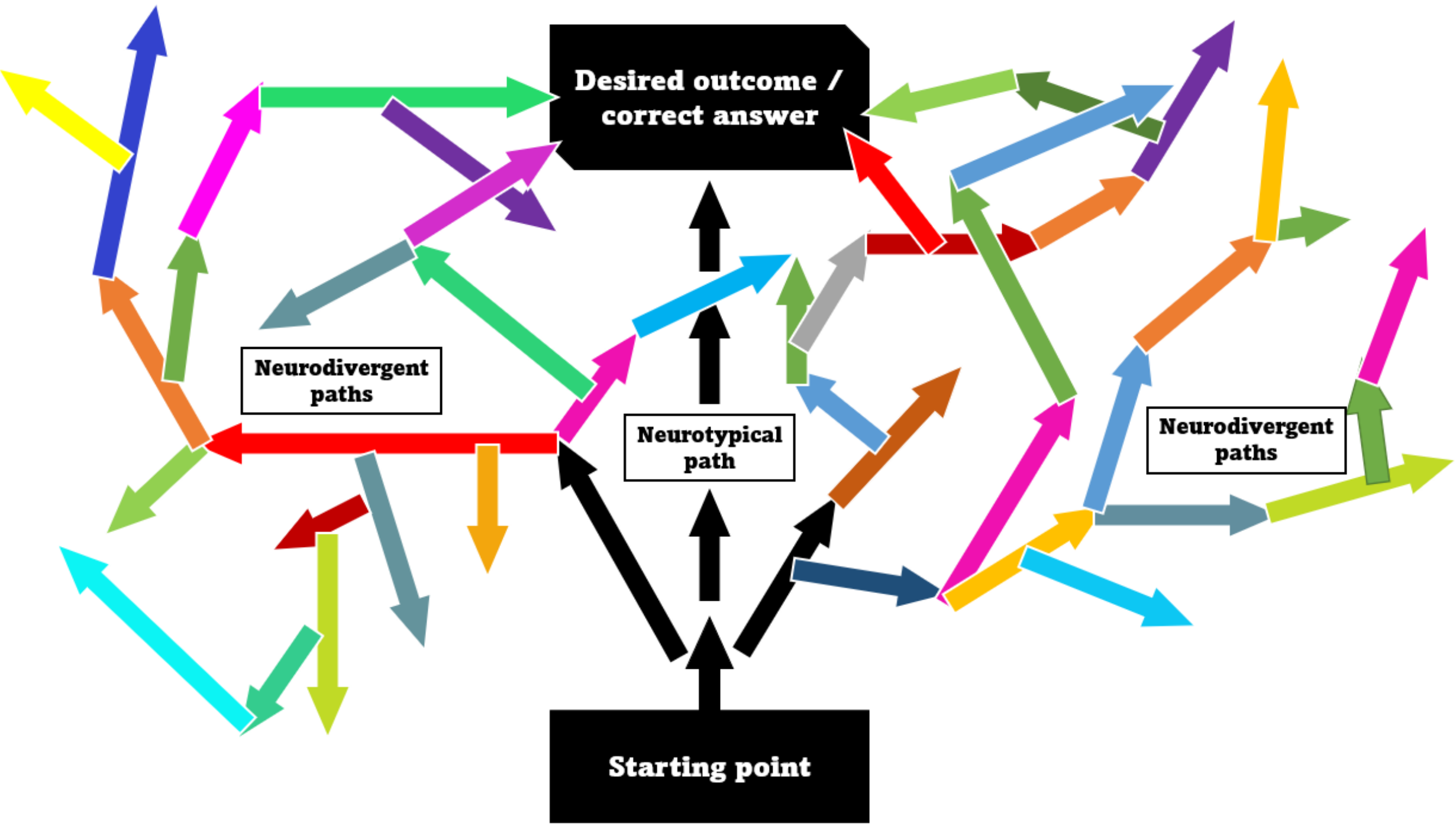

***Fig. 1.*** *Marsden's firework thought.*

The way I have framed this, it may seem like being neurodivergent and constructing a neuroqueer literacy creates more barriers than bridges in Marsden's physics career. However, Marsden was quick to correct this in our interview. Immediately after providing these comparisons between themself and their neurotypical peers, they explain that though their neurotypical peers have an easier time performing physics tasks, this neuroqueer thought process makes them "feel like I'm hot shit sometimes," because "there'll be some situation where the whole linear A to B thing doesn't work anymore, but because I had to spend all this time detangling for myself, it's easier." While textbook physics problems often require a linear A to B problem-solving strategy, research and reality are often not as linear. To use Marsden's words, "I'm capable of kind of breaking things down into smaller steps, because I feel like I'm forced to do that all the time. And so, when it comes down to the times where other people also need that, I feel like I can help." In thinking and performing non-linearly, Marsden gains an ability to empathize and help others navigate their own thought processes; something for which solving problems linearly does not leave much room.

**Tom (Undergraduate student, identifies with being autistic, depression, anxiety)**

Tom provided a very interesting perspective on how they construct their own neuroqueer literacy in physics and how it affected their experience in a specific class, quantum mechanics. Right out of the gate on the first question, "how do you engage in physics," they set themself apart from "typical" physicists, sharing that they think in "a more visual and abstract way." They specify, "I'm less grounded in the math, and more grounded in being able to visualize and understand what is going on."

In our interview, Tom indicated an understanding of neuroqueer thought, establishing their own "pet project," saying:

> This has been a little pet project of mine in my department where I've harassed people about like, 'How do you think when you do physics?' Because I think in a very, 'I see shapes, I have a sense of what is going on.' [But] I've met other people who … they literally have a whiteboard in their head that they do everything on, and they can't visualize anything.

While a whiteboard in one's head may sound visual, this is a case of Tom, a purely visual thinker, describing non-visual thought. This is instead meant to indicate that the person in question thinks either in 2-dimensional written word (unlike Tom's 3-d thought), or in a non-visual (e.g. numeric, formulaic) way. Tom only thinks in terms of three-dimensional pictures and shapes, and thus their understanding and conceptualization of the world is solely informed by this framework. All Tom's knowledge is constructed through this lens.

Knowledge construction grounded in visualization, Tom says, made understanding quantum mechanics a significant challenge. They state about quantum mechanics, "there is no visual analogy for what is going on and that was something that bothered me a lot." It bothered Tom to the point they asked their professor for a more Tom-specific problem-solving perspective. This professor could not give them that perspective they needed, "It's just the math. There is no more of an explanation," Tom describes their conversation. However, this disconnect did not deter Tom, as they see themself as a capable physicist, which may add some resilience to how they take on challenges. In fact, Tom says, this added challenge which came from the presentation of quantum mechanics as a purely mathematical and non-visualizable subject "that pushes me to keep trying to figure out how this works. How do I understand this in a sense, where I know, start to finish, what is going on?"

Tom also distinguishes their neurodivergent way of processing information from the more neurotypical style. "I- it depends on the subject, but I would generally say I don't study in a, like, [neurotypical way]. I don't open a textbook and stare at it for five hours," they say. "But I don't- … to make an analogy, instead of eating like 3 large meals a day, I eat a bunch of small meals throughout the course of the day, and that's kind of the way I process information… by working on it, a little bit, frequently." Not just how they perform physics, but how they study and learn is constructed differently from their neurotypical peers' methods.

Tom generalizes their experience, stating:

> Learning while being neurodivergent, has always kind of created some level of complication, because people try and lay out a path of like 'you should study this way. You should work in this way. You should problem in this- by these patterns.' And I never did that. I had to find my own way of doing that, which I found value in, but caused me to struggle a lot with, 'am I doing the right thing? Should I be doing things this way? Why aren't other people doing things this way? What is that? … What does that process look like?

In this, they share similarities with Marsden, who, in our interview, spoke a lot about how the path neurotypical educators lay out for neurodivergent students simply does not align with the orientation of their brains. In privileging the ways neurotypical people solve problems, educators alienate, cause complications, and disconnect their neurodivergent students from their class where they otherwise have strengths. This is especially concerning when neurodivergent students like Tom attribute much of the ways they think and the things which help them be better physicists outside the classroom to being neurodivergent, stating their being neurodivergent aids in "abstract problem solving, with dealing with, you know, frustrating situations with learning complicated and abstract material" and "bend[ing] myself socially and intellectually in ways in which other people cannot do." Educators do neurodivergent students like Tom a disservice when they construct their lessons around conventional, neurotypical methods to the exclusion of other methods, as neurodivergent learners construct knowledge unconventionally.

**Louis (Graduate student, identifies with BPD, anxiety)**

Louis did not speak much about his actual conception of physics per se, but instead spoke about his problem-solving methods and how neurotypical-normative assessment negatively impacts him in his field. "I like to work around the problem," he says, "whereas during exams I kind of feel like you don't have the time to go around the problem. You have to look at the problem, know exactly where you're trying to get to, and go there." He elaborates on what it means to "go around the problem," saying "I just try to understand the problem, like, really in depth." This correlates to the indirect and interconnected style of thinking described by both Tom and Marsden, albeit seemingly more cohesive and less fragmented than their thought process.

Louis conceptualizes physics and physics tasks in a nonlinear manner. Louis does not do physics head-on like his neurotypical peers; instead, his strengths lie in being able to fully understand everything about a system before attempting to provide a solution. This is a strength, Louis indicates, which is beneficial outside of the classroom. He says about being neurodivergent, "you can see things that I guess other people might have missed, or might not have looked at the same way… We just see the world through a different lens, which I think is very useful in the field of science."

The conflict between how physics ability is measured and how he does physics is not lost on Louis. When I asked him, "are there any systems in place which hinder you doing physics as a neurodivergent physicist?" Louis immediately responded, without hesitation, "God, exams, I hate exams." About exams, he says, "they're not a good reflection on how well I understand the material, and so that kind of hurts me academically, like grade-wise." Speaking about an exam he took prior to our conversation, he notes, "the exam problems we had were new problems, problems I'd, like, never seen before." He elaborates, "which I don't like on exams, because you're being tested on like essentially brand new material." This coincides with his assessment of his own problem solving. It takes time to learn and understand fully all parts of a problem, oftentimes more than a day (like Tom's multiple-snacks analogy to learning physics). When presented with a new problem, one which he has never seen before, he does not have time to analyze the intricacies of the problem to make it comprehendible within his neurodivergent thought processes.

Louis treats each problem as unique, not because he is unwilling to engage in linear problem solving, but because his mind is not oriented in that way. Figure 2 provides a visualization of Louis and other neurodivergent people "go around the problem," elucidating how useful this method of sense-making and problem-solving can be for neurodivergent people in the field, while requiring more effort than their neurotypical peers when dealing with linear tasks, like an exam. Neurodivergent learners do not perform physics the way their neurotypical peers perform physics. They instead perform tasks through their own embodied methods, whether examining every possible detail like Louis, or navigating "firework thought" like Marsden.

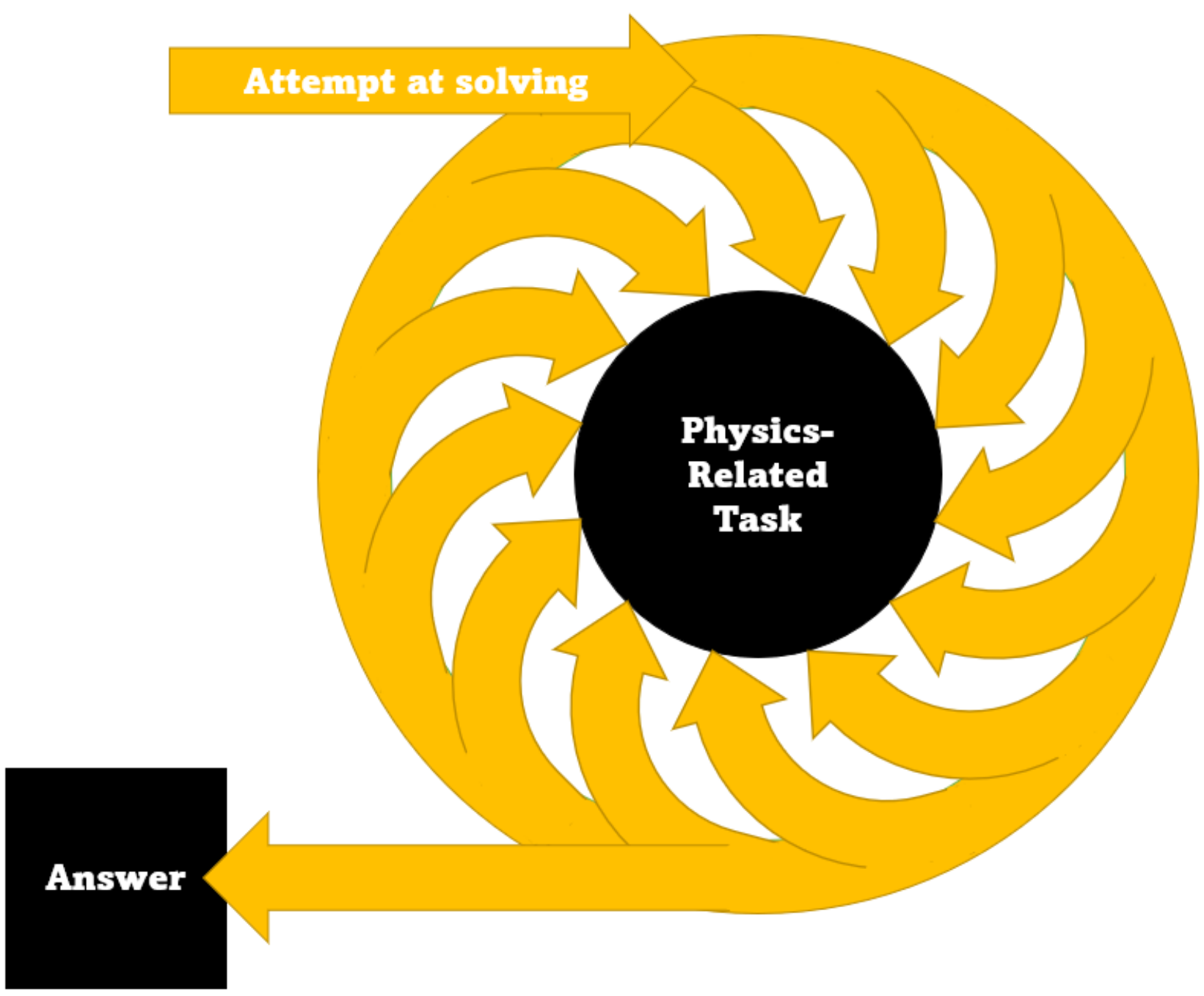

***Fig. 2.*** *Louis's "going around the problem" problem-solving process.*

**Elizabeth (Undergraduate student, identifies with being autistic, OCD, depression)**

"I can't visualize things in 3-D... I only visualize in 2D, which is very difficult when you're a physicist who wants to model things," Elizabeth describes how she thinks. She explains that she is applying to graduate school for condensed matter physics, a subject in which understanding the way matter is constructed and operates three dimensionally is critical. She tells me about the struggles she has faced due to instruction being ill-suited to the way her brain works: "[Modeling] was always a very, like, large challenge for me." She says, "[instruction methods] hindered a lot of the way that I understood things conceptually, because I feel like very often [professors] assume that you can [visualize in 3-D]." Upon asking for an example, she explains, "I was very confused when my professor would be like, 'so imagine an atom is coming in towards another atom, or there's a rocket flying.'" She continues, "I'm like I can't. [pause] I know what that means, but I can't physically see it in my head and that causes a lot of problems for me."

Elizabeth has had much support from her professors, friends, and herself regarding this didactic barrier. "Modeling in physics is so [important]," she reiterates, "so one thing that I've personally done is gotten really good at drawing very good diagrams." While physics and other science educators regularly emphasize the importance of drawing diagrams (Maries & Singh, 2017; Tang et al., 2019) in the hopes of helping students of all neurotypes, Elizabeth describes that the way we teach this modeling is still grounded in the assumption that everyone conceives of and perceives the world in the same way. "[Not visualizing in 3-D] causes a lot of problems for me, until I finally, like, told my professor [about it]." She explains how her professor reorients the material to match the orientation of her learning. "I was like, 'hey, I can't do this.' … [my professor] gave me really good advice of thinking about 3-D modeling in 2-D in the way, like, you do an engineering diagram that's broken apart." With this help, she explains, "I could do a kind of 3D model in a 2D form."

This speaks to a larger point Elizabeth makes in our interview. Namely, "I don't think physics can be done alone. It's too complicated. And you need multiple brains coming together to work on a problem." She self-reflects, saying, "I think for a long time I tried to brute force my way of doing it alone, and I came to a realization that it's just not possible." Neurodivergent students need educators to aid in their neuroqueer Literacy development, not reject it (and, by proxy, the neurodivergent thinker) in order to match convention.

**Albert (Graduate student, identifies with ADHD, OCD)**

"Once [my graduate cohort and I] started working together, and they saw, like, you know, I'm just seeing these connections in a different way," Albert tells me, clarifying a vignette about others' perception of him as a physicist. Albert provides me with an interesting perspective on how we define someone as "good" at a field which coincides with what has already been said, that being capable or scientifically/mathematically literate is relational (e.g. Skemp, 1978) and dependent on unseen processes which conform to our hegemonic expectations of success. He explains that others in his cohort of graduate students identified him as "good at math," which he quickly dismissed as false. He explained where he thinks this error comes from, saying that we "characterize [other people] as being innately good at things in a way that you don't want to see what they're actually doing… I'm just seeing these connections in a different way."

"I'm learning [physics] in a different way, and it's a less successful way [by hegemonic standards], and therefore it's a worse way," Albert tells me. "I was talking about how there's not a worst way [to learn], but like that's my intellectual knowledge as opposed to, like, my emotional thinking." In our interview, he indicates that while he, as a teacher, knows that there is no "true" or "correct" way to learn, this knowledge creates cognitive dissonance when applying it to himself. As the way his brain learns is not oriented in the same way as the way his classes are taught and assessed, despite knowing that there is no "wrong" way to learn, he still reactionarily believes himself to be worse at physics than his peers. The neurotypical hegemony with which we design our classes not only creates an external mis-defining of scientific literacy or competence, but also engenders an internalized ableism in observably competent practitioners. Despite concrete proof to the contrary, such as Albert's degree in physics and advancement to Ph.D. candidacy in physics, this belief that he has a "worse" way of learning persists in the back of his mind.

Understanding his neurodivergent identity has helped Albert develop strategies for success in his field. "I was thinking, like, I want to be better than these people," he says, explaining his unhealthy thought process before self-reflection on his own identity, "or I want to be the best at something." He explains that he has put effort toward reorienting his coping mechanisms to better align with success if his classes will not be oriented with his brain. "I'm trying to rephrase my goals. It's like 'I want to be doing this, and better than I was doing before.' … or 'I'm gonna be doing my best.'" While these coping mechanisms are useful and should be applauded and recognized, it is critical to note that Albert, and any other student, should not be forced to construct coping mechanisms to follow their passion. Science literacy is incredibly important, personally and societally, and forcing people to learn and perform contrary to the way their minds work leaves many otherwise capable and passionate people behind.

## Discussion and Implications

I write this paper not to claim I have the answers to what a neuroqueer physics literacy looks like, but to vehemently shake educators and researchers and scream "there is something cool and important here! Dig deeper!" Neurodivergent learners construct knowledge from their own divergent thinking, like Tom, who thinks strictly pictorally, or Marsden who solves problems through *firework thought*. There is no one way to think about physics, mathematics, or science in general, and educators do students and peers a disservice in privileging one way of knowing over another.

Neurodivergent physicists construct a non-normative, neuroqueer literacy in physics. Their minds are oriented counter to the neurotypical hegemony in education. When we construct our definition of scientific literacy from one perspective, we eliminate the potential of learning from different perspectives. By neuroqueering scientific literacy and encouraging non-neurotypical perspectives on science, we open our fields up to progress made in unorthodox, new, and interesting ways.

All five interviewees indicated that their success was fueled by being in community with others, especially research mentors and instructors. Science is a social act, done through interconnected and interdependent collaborations amongst people. When research networks open up and celebrate the diversity of perspectives and cognitions which are present in humanity, the results advance the field, and open up opportunities for us to see new and innovative ways to solve problems. We will only see the full benefit of this diversity of thought in science when researchers and educators celebrate neuroqueer ways of being scientifically literate, actively working to combat academic ableism.

It is critical to note just how much of an emphasis these interviewees placed on how mis-aligned traditional assessment tools are for these neurodivergent students. As evidenced by Louis, Tom, and Marsden, some neurodivergent students perceive neurotypical assessment strategies as rewarding rote memorization or linear A to B thinking. Assessment constructed counter to the ways they construct knowledge has had detrimental effects on participants' internal and external perceived competence and their own identity, through grades and self-reporting. We, scientists, decide who gets to be scientifically literate based on assessment (and other gatekeeping) designed from one subjective perspective, hegemonically deemed objective. If we expand what constitutes scientific literacy to include neuroqueer literacies, it can't just be empty words and promises. We still gatekeep science counterproductively if we don't change assessment and pedagogical methods to reflect a celebration of neuroqueer literacy. We cannot change assessment and pedagogical methods without listening to and embracing the neuroqueer standpoints of neurodivergent learners. As such, the following six recommendations arise out of the above narrativization, taking lessons from the perspectives of the 5 neurodivergent participants. This is not the end-all-be-all of recommendations, and I strongly encourage educators to experiment with their own ways of neuroqueering pedagogy in conversation and community with neurodivergent learners and peers.

1. **Use "no surprises" assessment**.
    a. Timed assessment privileges students whose thought favors linear A to B connections while marginalizing others. As seen with Marsden and Louis's narratives, neuroqueer thinking is often nonlinear. Louis specifically describes exams as surprising students with questions they have never seen before. We mainly see educators treat this way of thinking as a problem which can be solved by giving students extra time on exams. However, have you ever tried to think about a series of problems for three-hours straight with no breaks? Quite frankly, it sucks, and in material terms can become more akin to punishment than accomodation. Instead, here are 2 alternative ways "no surprises" exams might be constructed:
        i. Create exams with problems students have seen before, to test students' application of methods.
        ii. Give students extended exams which test students' creativity and problem solving through multiple days.
2. **Present strategies for conceptualizing scientific concepts beyond purely visual or mathematical models.**

a. Neurodivergent brains do not always mesh with how educators hegemonically present scientific concepts. Many neurodivergent people, like Tom and Elizabeth, experience aphantasia or hyperphantasia, the inability or sole ability to construct visual mental imagery (MacKisack at al., 2022). Beyond being established good practice for science education (Taljaard, 2016), it is doubly important for neurodivergent students that their educators practice multisensory pedagogical techniques when delivering information, i.e. learners need to be provided with representations of information in multiple sensory modes.

3. **Explicitly break concepts into fundamentals and reiterate first-principles thinking.**
   a. By breaking down examples and concepts into basic units, educators can guide students' "firework thinking" in class towards productive conclusions. Be proactive and consider how students may bifurcate logic. Create flowcharts for methods (an example from my class is included in Fig. 3).

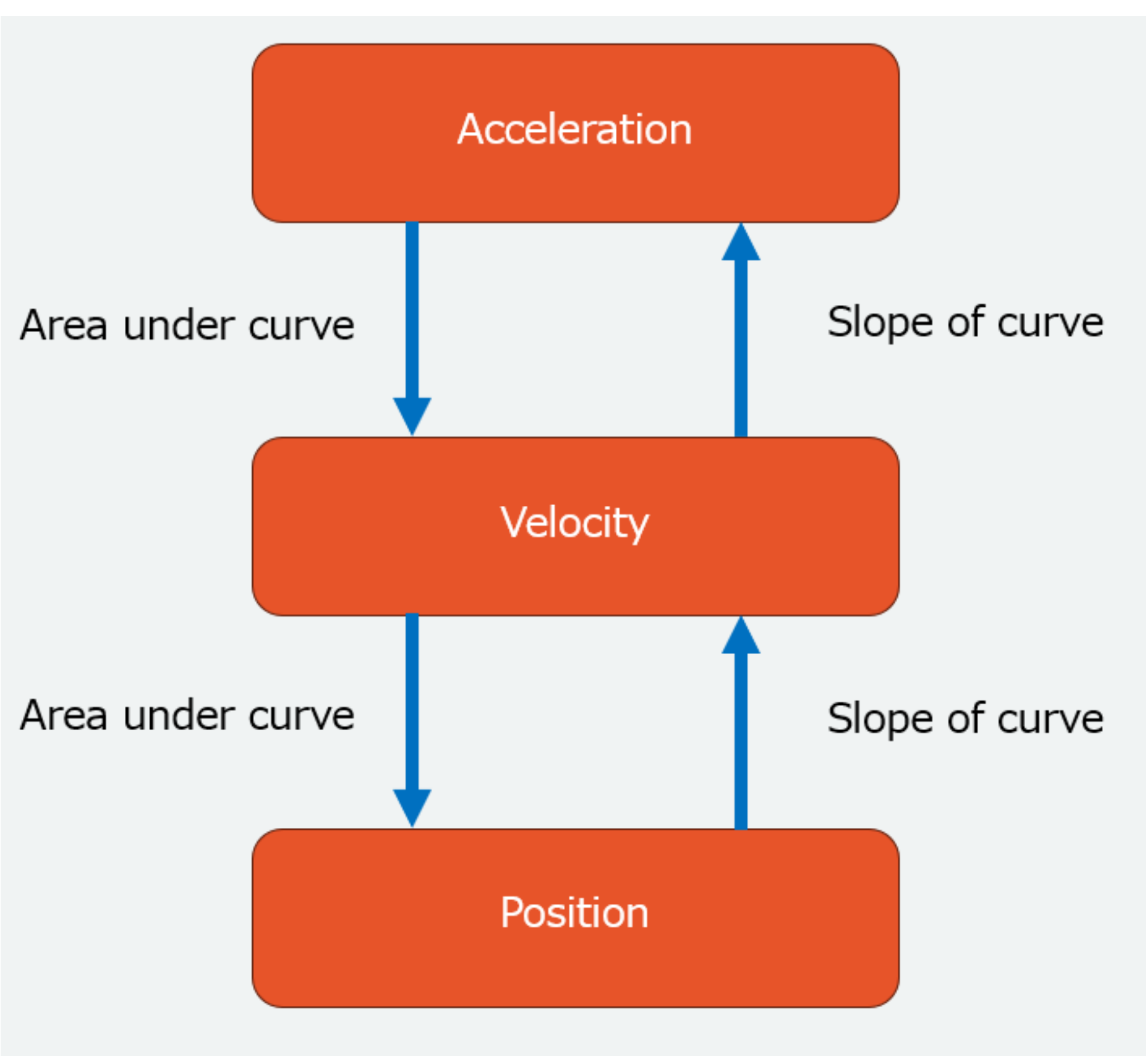

***Fig. 3:*** *A flowchart, which has had success in my introductory physics class, describing a method students can take to move from one kinematic concept to another mathematically.*

4. **Provide both top-down and bottom-up perspectives on concepts.**
   a. Neurodivergent people experience a variety of sense making and concept processing potentials. As an example, Louis describes "bottom-up thinking," processing all details before broad concepts. This is counter to the largely neurotypical "top-down" processing that educators construct their classes with. By using a mix of both top-down and bottom-up perspectives, we can engage all students. Instructors can incorporate bottom-up thinking in their lessons by, for example, building concepts from problem-solving perspectives, such as introducing the Lagrangian by providing systems where basic Newtonian physics cannot solve problems, requiring students (with aid from the educator) to build the Lagrangian from scratch.

5. **Provide lowest-stakes assessment.**
    a. Neurodivergent students report experiencing intense emotional, psychological, and sometimes even physical pain from receiving (or the threat of receiving) bad grades (Marsden cites this specifically). This anxiety can often result in assessment which otherwise would be used for self-reflection or motivation, to not be completed or send neurodivergent students into spiraling negative thoughts about their performance. Every person I interviewed, but especially Albert, Marsden, and Louis, indicated some experience with this profound level of anxiety. I recommend having ungraded assessment or assessment worth a miniscule percent of their grade, as opposed to assessment worth a significant amount of their grade (>5%), to help mitigate this.
6. **Reach out to neurodivergent learners (and pay them for their time).**
    a. Just reading this paper and using it to design a new curriculum would be inadequate, operating directly against the spirit of neuroqueering itself. While neurodivergent people have a lot in common, we are far from homogenous or generalizable. These recommendations are grounded in much scholarship and in the lived experiences of five neurodivergent physicists, but that means little when compared to the value of sincerely being in community with neurodivergent people. Ultimately, one must physically ask neurodivergent learners to review or test material in order to make sure their neurodivergent learning styles are included in class and valued commensurate with neurotypical learning styles. Instructors need to treat each specific student as a community partner and as an expert in their own needs. I recommend, as a tangible thing, that instructors make time to be available and to welcome discussions about student needs. It is also critical that educators do not tokenize these students and instead treat them as contractors providing a skillset educators do not possess—that is to say, by compensating them for their time (Oleynik et al., 2022).

In the spirit of neuroqueer literacies and the neurodivergent community, this work is not the end-all-be-all guide to developing neuroqueer literacies in science. Future research can and should expand upon this idea of neuroqueer literacies, expanding this dataset, developing this idea past the case of physics as a field, and understanding how to change science pedagogy for an equitable classroom, among many others.

To borrow another's words, my aim with this paper is "to resolve the apparent paradox whereby a community allegedly marked by sameness is in fact so broadly diverse and heterogeneous; in emphasizing choice and active engagement, it signifies members' experienced likeness not as an essential property, but as a collective achievement" (Belek, 2022, p. 630). We, both as scientists and as neurodivergent people, are not a homogenous set, yet the way we define scientific literacy assumes that we are homogenous and, even more egregiously, that we make sense of the world neurotypically. This contributes to systemic academic violence against neurodivergent and disabled learners, and those wounds can fundamentally alter and harm how neurodivergent people understand ourselves and our relationship with the world at large. As educators, we need to work together and make our classrooms equitable, to question, make strange, and subvert the ways neurotypical conventions are forced on our students as the singular "correct" way of being and knowing.

It is our duty to make the classroom we dream about the one we teach in now. We, largely, have already rejected the banking model (Freire, 2000) as aspirational or equitable, and acknowledged that students bring their own multifaceted and diverse experiences and intellectual resources into the classroom. It is time that we take another step forward and acknowledge that students bring their own multifaceted and diverse brains with them into the classroom. Good pedagogy should *both* leverage student resources to reinforce concepts *and* leverage the way students' minds absorb information and solve problems. Great pedagogy must encourage student growth and celebrate the queer and unique ways of doing science that students develop, opening the path that they might lead us to new, exciting, and uncharted territory.

## Limitations

All participants identified as white. People of Color have regularly been excluded from both physics (Cochran et al., 2021) and the neurodiversity community (Giwa Onaiwu, 2020), and their voices are once more missing from this conversation. It is therefore important that any reading of this work be cognizant that it is written from a white point of view and includes only perspectives from white neurodivergent physicists. Additionally, this research consists of a sample of five neurodivergent physicists. Though I believe that their stories are representative of a larger phenomenon of neuroqueered learning and performance, I report on only five perspectives of people who all share a common field. Future research should include perspectives from more people, comprising different fields, to build on this conversation.

## Afterword: Reflection on Intersubjectivity

How do you do research in a traditional way while calling for a departure from traditional ways of knowing and doing science? With much pushback and rejection, as it would turn out. This paper was originally a pet-project that developed organically during the data collection phase of my dissertation. My dissertation focused on how neurodivergent physicists develop an identity/identify as neurodivergent physicists (a distinction for which the lines are, personally, too blurry for me to draw in an afterword. Are we always in a state of becoming? Do we ever identify permanently?). However, it quickly became apparent that the neurodivergent physicists that I interviewed shared one major thing in common: they did and learned physics differently. Retrospectively, this fact may seem obvious: *neuro* – of the brain/mind, and *divergent* – departing from the norm, therefore *neurodivergent* – departing from the norm *vis-à-vis* "brain/mind stuff". However, for a neurodivergent physicist like myself, the fact that this phenomenon of "doing physics differently" was so palpable in the experiences of my interviewees was eye-opening. I, Dr. Liam G. McDermott, don't just personally do physics differently; neurodivergent people do physics differently! If only there was some paper which talked about that phenomenon in detail.

So, I went to work in my free time writing this paper. I hit a snag pretty early on in the process, however. None of my qualitative research textbooks laid out a formula for research that said "Hey! Everybody! There's something weird here! Something cool and weird is happening come look!" No matter, though, I know my Queer Theory and I know that good, queered work fucks with traditional methods. Someone in the peer review process will see how cool this idea is and show me how to queer my work to where it's audience friendly. However, while the peer reviewers in traditional journals did think the idea was worthwhile, the queered method of narrativizing interviews did not suit their expectations for this genre of writing. I was given two choices: un-queer the method in favor of traditional phenomenology/case study/narrative and publish, or put the paper away in a box for a future study. I felt this work needed to be read by educators, so I tore apart the paper to un-queer it enough for a more standard journal. Very fortunately (for my brain and for this paper), I was approached by the folks at JTM-ME with the express commitment that they wanted to keep all the neuroqueerness of the original document and expand it further.

The journey since agreeing to publish with JTM-ME has not only strengthened this work past its original bounds, but was incredibly affirming to both my neurodivergent identity and my neurodivergent professional practices. The review process involved meetings with Dr. David Bowers (they/them) who walked through the goals I had for this paper and how we could bring them to fruition in the final product. They let me stew on their recommendations for an extended period of time, no questions asked, and helped build the paper you see today. The peer review process of JTM-ME was exactly what I wish to see in a review process: full of excitement for the potential of research (with the checks and balances of peer review I've come to expect), rather than the antagonistic "reviewer 2" modality of review. Review was a supportive dialogue, not an adversarial debate. Because of this I was able to refine and build upon earlier forms or this paper, rather than scraping it to its bones and leaving it hollow. For that itself, I am deeply grateful.

## Acknowledgements

I would like to acknowledge and thank my graduate advisor, Dr. Geraldine Cochran, for encouraging me to take on this project on my own and for her discussions on power and positioning in research. I would also like to thank and acknowledge my current post-doc advisor, Dr. Erin Scanlon, for encouraging me to keep working on this paper early on in our research partnership.

Special thanks to Dr. David Bowers for encouraging me to "neuroqueer the fuck out of" this paper, and for seeing something special in this work. I'd also like to thank them for editing and reviewing this paper. Without their help, this would have never seen the light of day.

## References

Ballentine, K. L. (2019). Understanding racial differences in diagnosing ODD versus ADHD using critical race theory. *Families in Society*, *100*(3), 282-292.

Belek, B. (2023). 'A smaller mask': Freedom and authenticity in autistic space. *Culture, Medicine, and Psychiatry*, *47*, 626-646.

Bhattacharya, K. (2017). *Fundamentals of qualitative research: A practical guide.* Routledge/Taylor & Francis Group. https://doi.org/10.4324/9781315231747

Cochran, G.L., Hyater-Adams, S., Alvarado, C., Prescod-Weinstein, C., Daane, A.R. (2021). Social justice and physics education. In C. C. Ozaki & L. Parson (Eds.), *Teaching and learning for social justice and equity in higher education*. Palgrave Macmillan.

Freire, P. (2000). *Pedagogy of the oppressed* (30th anniversary ed.). Continuum. (Original work published in 1970)

Giwa Onaiwu, M. (2020). I, too, sing neurodiversity. *Ought: The Journal of Autistic Culture*, *2*(1), 58-67.

Hadden, S. (2023). *How do neurodivergent people engage with tone in digital spaces? A study on the written expression of tone and intent.* [Master's thesis, Arizona State University]. KEEP.

Heasman, B., & Gillespie, A. (2019). Neurodivergent intersubjectivity: Distinctive features of how autistic people create shared understanding. *Autism*, *23*(4), 910-921.

Holmes, N. G., Olsen, J., Thomas, J. L., & Wieman, C. E. (2017). Value added or misattributed? A multi-institution study on the educational benefit of labs for reinforcing physics content. *Physical Review Physics Education Research, 13*(1), 1-12.

Jack, J. (2022). The cognitive vernacular as normative mandate in habits of mind. *College English*, *84*(4), 335-355.

Johnson, M. L., & McRuer, R. (2014). Cripistemologies: Introduction. *Journal of Literary & Cultural Disability Studies, 8*(2), 127-147. https://www.muse.jhu.edu/article/548847.

Jones, S. D., Jones, M. W., Koldewyn, K., & Westermann, G. (2023). Rational inattention: A new theory of neurodivergent information seeking. *Developmental Science, 27*(4), 1-19.

Jurgens, A. (2020). Neurodiversity in a neurotypical world: An enactive framework for investigating autism and social institutions. In H Rosqvist, N. Chown, & A. Stenning (Eds.), *Neurodiversity studies* (pp. 73-88). Routledge.

Kleekamp, M. C. (2020). "No! turn the pages!": Repositioning neuroqueer literacies. *Journal of Literacy Research*, *52*(2), 113–135.

Lang, M. (2019). Seeing in color: How are teachers perceiving our diverse autistic students?. *Ought: The Journal of Autistic Culture*, *1*(1), 80-92.

MacKisack, M., Aldworth, S., Macpherson, F., Onians, J., Winlove, C., & Zeman, A. (2022). Plural imagination: Diversity in mind and making. *Art Journal*, *81*(3), 70–87.

Maries, A., & Singh, C. (2017). Do students benefit from drawing productive diagrams themselves while solving introductory physics problems?: The case of two electrostatics problems. *European Journal of Physics*, *39*(1), 015703.

McDermott, C. (2022). Theorising the neurotypical gaze: Autistic love and relationships in the bridge (Bron/Broen 2011–2018). *Medical Humanities*, 48(1), 51–62. https://doi.org/10.1136/medhum-2020-011906

McRuer, R. (2004). Composing bodies; or, de-composition: Queer theory, disability studies, and alternative corporealities. *JAC*, *24*(1), 47–78. http://www.jstor.org/stable/20866612

Milton D. (2012). On the ontological status of autism: The 'double empathy problem'. *Disability and Society*, 27(3), 883–887.

Odden, T. O., Lockwood, E., & Caballero, M. D. (2019). Physics computational literacy: An exploratory case study using computational essays. *Physical Review Physics Education Research, 15*(2), 020152. https://doi.org/10.1103/physrevphyseducres.15.020152

Oleynik, D., Scanlon, E., & Chini, J. (2022, July 13-14). The epic and the tragedy: Narratives of a disabled physics student. Paper presented at Physics Education Research Conference 2022, Grand Rapids, MI.

Radulski, E. M. (2022). Conceptualising autistic masking, camouflaging, and neurotypical privilege: Towards a minority group model of neurodiversity. *Human Development*, *66*(2), 113–127. https://doi.org/10.1159/000524122

Rodriguez Espinosa, P., Pichayayothin, N. B., Suavansri, P., French, J. J., Areekit, P., Nilchantuk, C., Jones, T. S., Mam, E., Moore, J. B., & Heaney, C. A. (2022). Found in translation:

Reflections and lessons for qualitative research collaborations across language and culture. *International Journal of Qualitative Methods,* 21, 1-12. 16094069221101280.

Nicola Rollock (2012) The invisibility of race: Intersectional reflections on the liminal space of alterity. *Race Ethnicity and Education, 15*(1), 65-84. 10.1080/13613324.2012.638864

Skemp, R. (1978). Relational understanding and instrumental understanding. *The Arithmetic Teacher, 26*(3), 9-15.

Smilges, J. L. (2021). Neuroqueer literacies: Or, against able-reading. *College Composition and Communication*, *73*(1), 103-125.

Tang, K. S., Won, M., & Treagust, D. (2019). Analytical framework for student-generated drawings. *International Journal of Science Education*, *41*(16), 2296–2322. doi:10.1080/09500693.2019.1672906

Taljaard, J. (2016). A review of multi-sensory technologies in a science, technology, engineering, arts and mathematics (STEAM) classroom. *Journal of learning Design*, 9(2), 46-55.

von Below, R., Spaeth, E., & Horlin, C. (2021). Autism in higher education: Dissonance between educators' perceived knowledge and reported teaching behaviour. *International Journal of Inclusive Education*, *28*(6), 940-957.

Walker, N. (2014). *Neurodiversity: Some basic terms & definitions*. Neuroqueer. https://neuroqueer.com/neurodiversity-terms-and-definitions/

Walker, N. (2021). *Neuroqueer heresies: Notes on the neurodiversity paradigm, autistic empowerment, and postnormal possibilities*. Autonomous Press.

Wilson, J. D. (2017). Reimagining disability and inclusive education through universal design for learning. *Disability Studies Quarterly*, *37*(2).

Yuliati, L., Parno, Hapsari, A. A., Nurhidayah, F., & Halim, L. (2018). Building scientific literacy and physics problem solving skills through inquiry-based learning for STEM education. *Journal of Physics: Conference Series*, 1108, 012026.